# Hierarchical Trait-State Model for Decoding Dyadic Social Interactions

Qianying Wu[1], Shigeki Nakauchi[2,3], Mohammad Shehata[3,4,*], Shinsuke Shimojo[4]

[1]Division of Humanities and Social Sciences, California Institute of Technology, Pasadena, CA, 91106
[2]Department of Computer Science and Engineering, Toyohashi University of Technology, Toyohashi, Japan, 441-8122
[3]The Institute for Research on Next-generation Semiconductor and Sensing Science (IRES2), Toyohashi University of Technology, Toyohashi, Japan, 441-8122
[4]Division of Biology and Biological Engineering, California Institute of Technology, Pasadena, CA, 91106
* Correspondence should be addressed to Mohammad Shehata (mohammad.shehata@gmail.com)

## Author Contributions
QW, MS, and SS Designed Research; QW and MS Performed Research; QW, MS, SS, and SN Analyzed Data and Wrote the paper.

## Abstract

Traits are patterns of brain signals and behaviors that are stable over time but differ across individuals, whereas states are phasic patterns that vary over time, are influenced by the environment, yet oscillate around the traits. The quality of a social interaction depends on the traits and states of the interacting agents. However, it remains unclear how to decipher both traits and states from the same set of brain signals.

To explore the hidden neural traits and states in relation to the behavioral ones during social interactions, we developed a pipeline to extract latent dimensions of the brain from electroencephalogram (EEG) data collected during a team flow task. Our pipeline involved two stages of dimensionality reduction: first, non-negative matrix factorization (NMF), followed by linear discriminant analysis (LDA). This pipeline resulted in an interpretable, seven-dimensional EEG latent space that revealed a trait-state hierarchical structure, with macro-segregation capturing neural traits and micro-segregation capturing neural states. Out of the seven latent dimensions, we found that three that significantly contributed to variations across individuals and task states. Using representational similarity analysis, we mapped the EEG latent space to a skill-cognition space, establishing a connection between hidden neural signatures and social interaction behaviors. Our method demonstrates the feasibility of representing both traits and states within a single model that correlates with changes in social behavior.

## Significance Statement

This study presents a novel computational approach to model the neural representations of both traits and states during dyadic interactions. Using high-dimensional EEG data from a team flow task, we developed a pipeline that classifies traits and states within a low-dimensional latent neural space. This space exhibits a hierarchical trait-state architecture, is mathematically interpretable, and strongly linked to observable social behaviors. Our method offers new insights into the brain's role in social dynamics and provides a versatile tool for neuroscience and psychology.

# Introduction

      Neuroscientists have been studying the neural underpinnings of social interactions for decades, aiming to improve social skills, facilitate socialization, and enhance team performance (Porcelli et al., 2019; Redcay and Schilbach, 2019). The neural representations during social interactions depend on two factors: the social context and the innate characteristics of the interacting individuals (Finn et al., 2018; Guthrie et al., 2022). The former refers to the individuals' flexible, context-dependent states, while the latter represents more fixed, inherent traits. For example, consider the interaction between two individuals. Both of their brains might exhibit similar sensitivities in responding to incoming social cues (e.g., gestures, facial expressions, etc.) during an interaction, a stable trait; as a result, they simultaneously experience pleasurable socialization, a temporary state. In contrast, the interaction between individuals with different sensitivities to social cues might result in less pleasurable socialization, another temporary state. Various psychological theories compare traits and states: traits are patterns of the brain signals and behaviors that are stable over time but differ across individuals, whereas states are phasic patterns that vary over time, are influenced by the environment, yet oscillate around the traits (Fleeson and Jayawickreme, 2015; Steyer et al., 2015; Tamir and Thornton, 2018).

      Empirical studies have shown numerous associations between neural activity patterns and traits and states (Chen et al., 2022; Dubois et al., 2018; Haynes and Rees, 2006). Similar to psychological traits and states, neurologically, brain activities also exhibit trait- and state-like features: the electrophysiological activities across the entire brain form a high-dimensional representation that differentiates among individuals (i.e., a trait like feature); from time to time, the execution of certain tasks trigger temporal changes to the activities that differentiates among states (i.e., a state like feature). It is important to clarify that the neural traits described in this study using EEG should not be confused with the descriptive personality traits mentioned in the psychology literature (e.g., Big Five personality (Corr and Matthews, 2020; John and Robins, 2021).

      Despite substantial efforts to characterize neural traits or states using brain imaging techniques, particularly electroencephalogram (EEG), previous studies have not represented both traits and states in the same model (Edla et al., 2018; Jalaly Bidgoly et al., 2020; Lin et al., 2010; Rashid et al., 2020; Rosli et al., 2021; Supratak et al., 2017). Therefore, a primary goal of the current study is to build an interpretable model to identify a parsimonious set of latent dimensions representing both traits and states in the brain. The second goal is to establish a connection between neural and psychological traits and states – demonstrating that variability in the neural representation of traits and states is associated with behavioral changes across different individuals (traits) and conditions (states). Here, we analyzed a hyperscanning EEG dataset collected in a previous study in which dyadic teams played a music rhythm game—a naturalistic social task—designed to induce three types of states: team only, flow only, and team flow (Shehata et al., 2021). Team flow is a unique state of socialization where two (or more) individuals collaborate on tasks with a shared purpose and strong commitment (van den Hout et al., 2018). It produces exceptionally positive subjective experiences and is thus distinct from individual flow

experience or ordinary team interactions, both psychologically and physiologically (Pels et al., 2018).

To achieve our goals, we developed a two-stage dimensionality reduction pipeline utilizing both non-negative matrix factorization (NMF) and linear discriminant analysis (LDA). NMF is a widely used dimensionality reduction technique in neuroscience research, capable of extracting low-dimensional, and meaningful features from high dimensional data (Liu et al., 2004; Xu et al., 2022). LDA also serves as a dimensionality reduction tool facilitating the classification of certain variables in the data, and has been successfully applied to identify latent traits of a rich behavior repertoire in mice (Forkosh et al., 2019; Serra et al., 2021). Using this pipeline, we reduced the whole-brain EEG data from 512 dimensions to a seven-dimensional latent EEG space and identified a trait-state hierarchy. This hierarchy was reflected through the macro- and micro-segregation patterns of the latent space data distribution. Next, we quantified the unique contributions of the latent dimensions (LD) in characterizing inter-individual and inter-state differences, and provided visualizations for biological interpretations of the LDs. Finally, we confirmed that the traits and states in the identified latent EEG space shared a similarity structure with a skill-cognition space, thus establishing a brain-behavior association.

## Results

**An interpretable latent EEG space captured inter-individual and inter-state variabilities**

In the current analysis, we utilized the behavioral and EEG data collected in a previous hyperscanning study (Shehata et al., 2021). In the study, 15 participants were matched to form 10 pairs (i.e., 5 participants were paired twice) based on their skill level, song preference, and team preference. Each pair of participants then played a music rhythm game in which both players tapped an iPad screen when visual cues (notes) reached a yellow judgment line (Figure 1A). Participants played under three task conditions: Team Flow, Flow Only, and Team Only. In the Team Flow condition, participants could see their partner's positive feedback while playing the songs. In the Flow Only condition, team interaction was disrupted by hiding the partner's positive feedback using an occlusion cardboard between the participants. In the Team Only condition, the flow state was disrupted by scrambling the music. Each team completed 6 Songs under each of the three conditions in a pseudorandomized manner for a total of 18 trials (Figure 1B). After each trial, participants rated their subjective experience on several dimensions, and the ratings were averaged to form the Flow, Team, and Team Flow indices. High-density (128 channels) hyperscanning EEG data were collected from both players while they played the game. The ratings and the EEG data confirmed that the three conditions induced three different states in the participants: compared to the other conditions, participants reported higher Team Flow Index, and exhibited enhanced interbrain synchrony in the Team Flow condition; during the Team Only condition, the Flow Index decreased; and during the Flow Only condition, the Team Index decreased.

Using the EEG data, we built a pipeline to achieve our first goal - identifying latent dimensions that represent inter-individual and inter-state variabilities through each participant's (i.e., intra-brain) EEG profiles. This pipeline consisted only of linear transformations to enhance model interpretability (Figure 1C). We started by calculating each participant's normalized power spectral density (PSD) over all the 128 EEG channels and 4 frequency bands (i.e., theta: 4-7 Hz, alpha: 8-12 Hz, beta: 13-30 Hz, and low-gamma: 31-50 Hz) as input features (i.e., 512 features). Next, to avoid overfitting (otherwise the number of features exceeds the number of samples – check methods for more details), we performed an initial dimensionality reduction using non-negative matrix factorization (NMF) and reduced the dimensions to 160, including the top 40 NMF components from each frequency band. Finally, we conducted linear discriminant analysis (LDA) on our selected NMF components to form a latent space that maximized the accuracy of classifying each sample's participation identity (i.e., a single experimental session, if a participant paired twice with different people, that counted as two participations) and task condition (e.g., participant 1's first participation under Team Only condition). Meanwhile, this LDA model served as a secondary dimensionality reduction and gave us a final latent space of 7 dimensions. During both stages of dimensionality reductions, the model hyperparameters (i.e., the number of NMF components and the number of LDA latent dimensions) were determined as a trade-off between the model complexity and the final prediction accuracy out of a 5-fold cross-validation (Figure 2A-C, see details in Materials and Methods). Our current choice of features resulted in a validation accuracy of 44.95% (chance level = 1.67%).

To further demonstrate that the ability to classify participations and states is due to the inherent features of our data (i.e., the existence of trait and state characteristics in the EEG profile) instead of an overfitting of noise, we conducted three levels of permutation tests: randomly shuffling the labels of both participation identities and task states, randomly shuffling the task state labels while preserving the participation labels, and randomly shuffling the participation labels while preserving the task state labels. If our classification was an overfitting of noise, we would likely achieve the same level of accuracy even after permuting the labels, indicating no valuable information in the true labels, thus the permutation helped us establish a null distribution of validation accuracy. However, our resulting validation accuracy was significantly higher than all three permutation-based null distributions (shuffle participation and state labels: $chance\ validation\ accuracy = 1.52\% \pm 0.76\%, p < 0.001$; shuffle state labels only: $chance\ validation\ accuracy = 31.74\% \pm 2.72, p < 0.001$; shuffle participation labels only: $chance\ validation\ accuracy = 1.74\% \pm 0.82, p < 0.001$, Extended Data Figure 2-1).

Because our goal is to better understand the data through a latent representation approach, we next applied our established pipeline to the entire dataset (i.e., 40 NMF components per channel, 7 latent dimensions, without training-validation partition) and examined several properties of this space. Our pipeline generated seven latent dimensions that explained 76.6% between- vs. within-class variance (Extended Data Figure 2-2A) and achieved an in-sample classification accuracy over 99% (Extended Data Figure 2-2B). We visualized this latent space by projecting each participant's trial-level intra-brain EEG data onto the top 3 latent dimensions (LDs) (Figure 2D) and found two levels of clustering: a major clustering (or macro-segregation) of participations (indicated by color) and a minor clustering (or micro-segregation) of task conditions within each participation (indicated by shape).

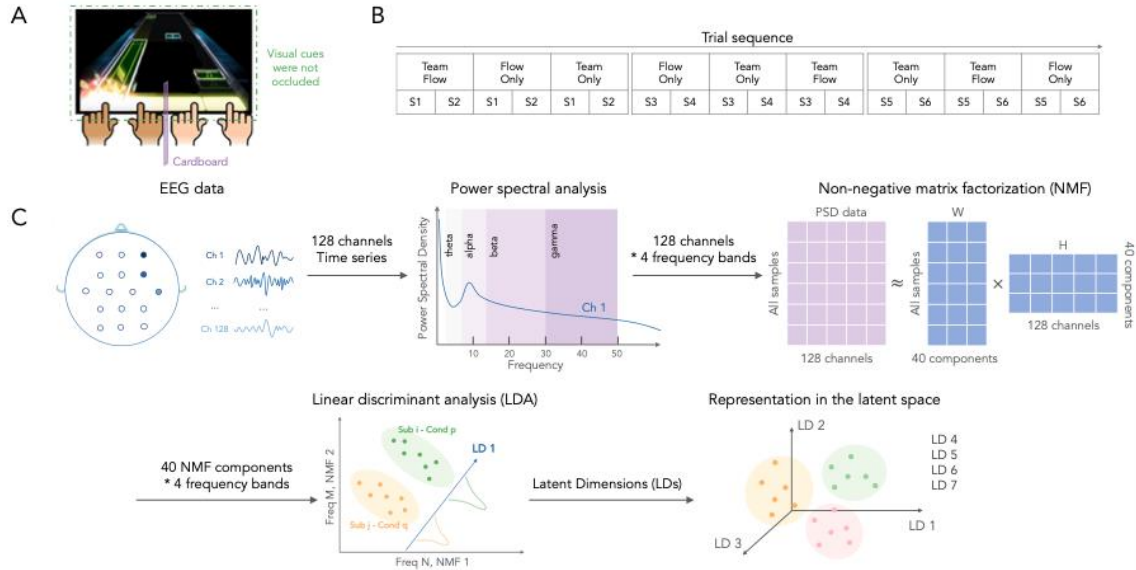

**Figure 1. Data acquisition and analysis pipeline.** (A) Illustration of the task. Two participants played the music game as a team, in which they both needed to tap an iPad screen when visual cues (notes) reached the yellow judgment line. In the Team Flow condition, the music was typical, and participants sat adjacent to each other. In the Team Only condition, the music was scrambled. In the Flow Only condition, cardboard separated the participants so they could not see each other's feedback. (B) The trial sequence: each pair of participants played 18 trials – six songs (S1-S6) under the three conditions in the pseudorandomized order presented. (C) The two-stage dimensionality reduction pipeline. Preprocessed EEG time series of 128 channels went through the power spectral analysis, resulting in the power spectral density (PSD) of each channel in the theta (4-7 Hz), alpha (8-12 Hz), beta (13-30 Hz), and low-gamma (31-50 Hz) frequency bands. Next, for dimensionality reduction, 40 latent components (out of 128 channels) were identified using non-negative matrix factorization (NMF) for each frequency band. Finally, all these NMF components were transformed into a 7-dimensional latent space through linear discriminant analysis (LDA) to maximize the distances between the participation x condition clusters.

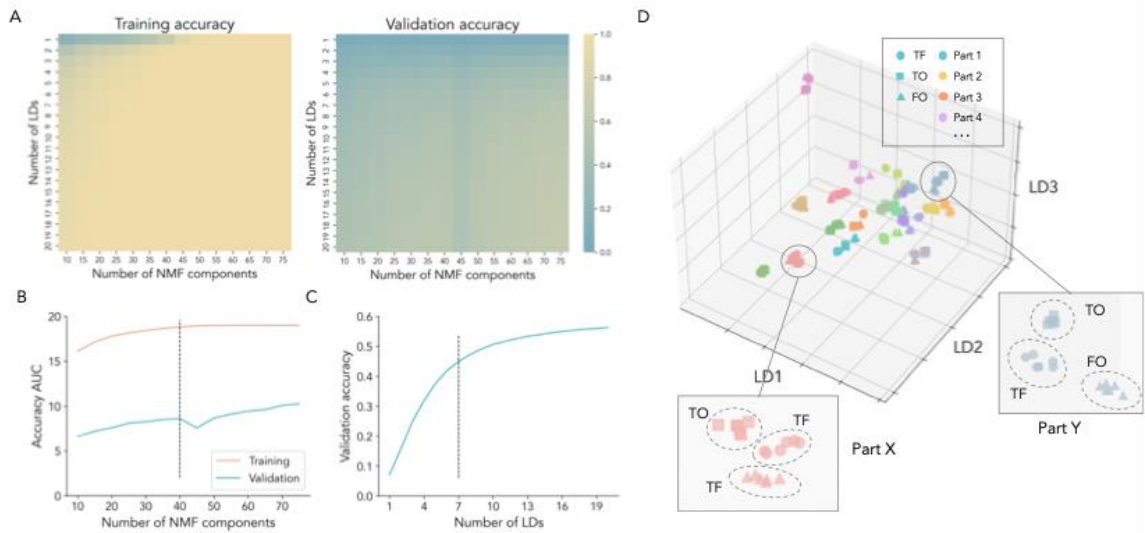

**Figure 2. Latent space representation of the team flow data shows macro- and micro-segregation.** (A-C) Feature selection of the two-stage dimensionality reduction. (A) Classification accuracy. Training (left) and testing (right) accuracies on the classification of each participation x condition class. Heatmap shows how the accuracy was affected by the choice of the number of NMF components (horizontal) and LDA latent dimensions (vertical). See a control analysis validating the significance of the prediction accuracy against permutation-based null distributions in Extended Data Figure 2-1. (B) Selection of NMF components. The amount of NMF components to include in the model depends on the area-under-curve (AUC) value of the classification accuracy on the testing set, across 20 LDA latent dimensions. (C) Selection of LDA latent dimensions. The amount of LDA latent dimensions to include in the model was decided as the knee point of the accuracy curve. (D) Visualization of the latent space and the macro & micro-segregation model. Representation in the top 3 latent dimensions. Each scatter represents the data of one trial after various stages of transformation. Color denotes the participation identity, and three shapes correspond to the three task conditions (Circle, TF = Team Flow; Square, TO = Team Only; triangle, FO = Flow Only). Two representative participations are zoomed in to show the macro-segregation between the participations and the micro-segregation between the different task conditions in the current 3D latent EEG space. For several in-sample LDA metrics, see Extended Data Figure 2-2.

**Macro- and micro-segregation of the latent EEG space correspond to an individual's traits and states.**

So far, the intra-brain latent EEG space showed a pattern of macro-segregation across participations (or individuals) and micro-segregation across task conditions (within each participation). We hypothesized that the macro- and micro-segregations represent the neural manifestations of traits and states, respectively, in a trait-state hierarchical structure. More specifically, each participant occupied a locus in the latent space corresponding to the personal traits (i.e., macro-segregation) due to unique EEG characteristics. These characteristics were subject to subtle variations around the loci depending on the external environments (e.g., experimental manipulation), and thus these variations represented an individual's state (i.e., micro-segregation). However, several alternative possibilities exist. In this section, we tested our hypothesis against these alternatives.

First, we wanted to test whether the macro- and micro-segregations are truly hierarchical. LDA does not guarantee a certain structure of the space, and there can be three possible structures. One possibility is that there is no hierarchy at all – any participation-state cluster can be close to or far from any other cluster (no hierarchy, Figure 3A). Another possibility is that states constitute macro-segregations, and traits constitute micro-segregations within each state (state-trait hierarchy, Figure 3B). This is plausible as a previous study has found several neural correlates that showed significant differences between the three task conditions, and within each condition there were smaller variabilities across participants (Shehata et al., 2021). The last possibility is our hypothesized model, in which traits form macro-segregations and states form micro-segregations within each trait (trait-state hierarchy, Figure 3C). In other words, individuals' EEG profiles differ largely from each other, and the variations across different task conditions are more subtle.

To distinguish these three possibilities empirically, we derived the inter/intra individual distance ratio as a testing metric. We first identified the centroids of each data class (i.e., all the samples that share the same participation and task state label) in the 7-dimensional latent space. Then, we defined the inter-individual distance as the Euclidean distance between any two participations for a given task condition, and the intra-individual distance as the Euclidean distance between any two task conditions for a given participation. In the case of state-trait hierarchy, intra-individual distances should be larger than inter-individual distances, so the ratio should be smaller than 1. In the case of no hierarchy, the inter/intra individual distance ratio should be equal to 1. Similarly, in the case of trait-state hierarchy, that ratio should be greater than 1. The actual inter/intra individual distance ratio from the data was 7.08, and was significantly greater than the null distribution ($null\ distribution: M = 1.00, SD = 0.048,\ p < 0.001$, Figure 3D). These results supported our hypothesis that the trait-state hierarchical structure is valid beyond visual appearance.

Next, while we assumed that the macro-segregations represent traits, a counterargument is that they instead represent features related to experimental setup during each instance of participation. If our hypothesis is correct, samples from the participants who participated twice in the study (labeled as two separate participations in the model) should locate closer to each other in the latent EEG space than samples from entirely different participants, since their innate neural features (i.e., traits) should be relatively stable even during different experiment participations. A one-tailed independent sample t-test showed that the Euclidean distances between two participations of the same individual ($M = 2.63, SD = 0.56$) are significantly smaller than the distances between any different pairs of individuals ($M = 3.74, SD = 0.90;\ T(188) = 2.75, p = 0.007$). In addition, we performed a hierarchical clustering across participations. Participants were labeled as adjacent if they occur within the first and second linkage steps. We found that 3 out of 5 repeated participations from the same participant were clustered adjacent to each other in the dendrogram (Figure 3E). These analyses suggests that the macro-segregation represent neural traits rather than random features associated with participation circumstances.

Thus far, we have successfully excluded several alternative possibilities. All the above analyses supported our trait-state hierarchy hypothesis, that the latent EEG space represents individual neural traits in the form of macro-segregations at a higher hierarchy, as well as neural correlates of the task states in the form of micro-segregations at a lower hierarchy.

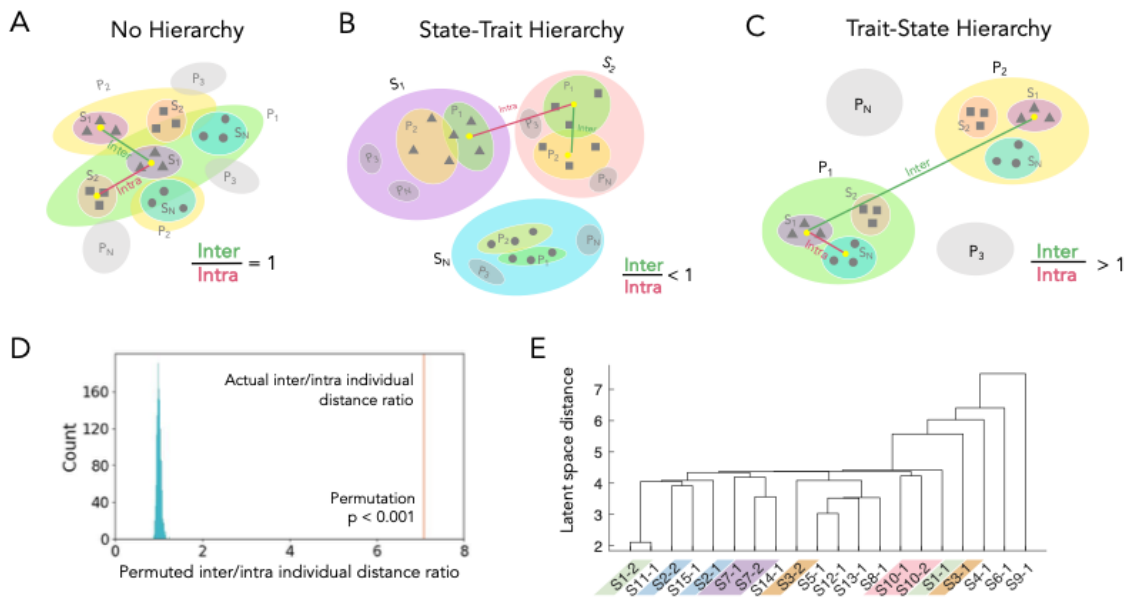

**Figure 3. Hypothetical models about the intra-brain latent space hierarchy.** (A) No hierarchy model. Inter- and intra-individual distances are of the same magnitude. (B) State-trait hierarchy model. State represents the macro-segregation, and trait represents the micro-segregation. Inter-individual distance is smaller than intra-individual distance. (C) Trait-state hierarchy model. Trait represents the macro-segregation, and state represents the micro-segregation. Inter-individual distance is greater than intra-individual distance. (D) Quantification of the segregation. The inter-individual distance is larger than the intra-individual distance, and their difference is significantly above chance. The vertical line is the inter-intra individual distance ratio from actual data, and the histogram shows the null distribution, calculated through 5000 permutations (i.e., shuffle the subject and task labels). (E) Hierarchical clustering of the task participation. Subject task participations are clustered based on their average latent EEG distance. Subjects who participated twice are highlighted. The repetitive participations are clustered closer to each other, indicating subject-specific, trait-like characteristics of the latent EEG dimension.

**Contributions of latent dimensions to inter-individual and inter-state differences**

We next investigated the contribution of each latent dimension to the characterization of trait and state variabilities, that is, which latent dimensions captured differences across individuals, and which latent dimensions captured differences across the three task states.

To identify LDs that captured inter-individual differences, we utilized the repeated participations in the dataset. We defined a repeated vs. non-repeated participation distance ratio to quantify the distance between the repeated participations of the same individual relative to the distance between any two different individuals. If an LD captures inter-individual differences, it should feature a larger non-repeated participation distance and a smaller repeated participation distance, thus the ratio is small. Among the seven LDs, LD1 ($M = 0.732, SD = 1.217, T(1923) = 8.256, p < 0.001$), LD3 ($M = 0.895, SD = 1.593, T(1923) = 6.296, p < 0.001$), and LD5 ($M = 1.158, SD = 1.541, T(1923) = 3.808, p = 0.001$) had significantly smaller repeated vs. non-repeated participation distance ratios compared to a bootstrapped null distribution ($M = 1.552, SD = 2.784$) indicating their contribution to the characterization of individual traits (Figure 4A).

To identify LDs that captured inter-state differences, we first normalized the LDs within each trial, and then performed one-way ANOVA across three conditions for each LD (Figure 4B). A significant main effect of condition was found in LD1 ($F(2,345) = 16.40, p < 0.001$), LD3 ($F(2,345) = 6.03, p = 0.019$), and LD5 ($F(2,345) = 7.42, p = 0.005$, $all\ p\ values\ after\ Bonferroni\ correction$). Post-hoc pair-wise Tukey tests showed that in LD1, all three conditions were significantly different from each other ($M_{TF} = -1.60, SD_{TF} = 4.73\ ;\ M_{TO} = -0.12, SD_{TO} = 4.87\ ; M_{FO} = 1.72, SD_{FO} = 3.57\ ; TF\ vs.\ TO: p = 0.029, TF\ vs. FO: p = 0.004,\ TO\ vs. FO: p < 0.001$). In LD3, Team Flow ($M_{TF} = 1.04, SD_{TF} = 4.76$) was significantly different from the other two conditions ($TF\ vs.TO: p = 0.015, TF\ vs. FO: p = 0.004$), whereas there was no significant difference between the Team Only ($M_{TO} = -0.41, SD_{TO} = 3.28$) and Flow Only condition ($M_{FO} = -0.63, SD_{TO} = 3.05$). In LD5, Flow Only ($M_{FO} = -0.91, SD_{TO} = 3.42$) was significantly different from the other two conditions ($FO\ vs. TF: p < 0.001, FO\ vs. TO: p = 0.038$) and the difference between the Team Flow ($M_{TF} = 0.75, SD_{TF} = 2.61$) and Team Only condition ($M_{TO} = 0.16, SD_{TO} = 2.16$) was not significant. Note that the sign of the LD values is not meaningful; thus we only focus on the statistical difference between the conditions. The above results suggested that LD1 is associated with a task component that differs across all conditions,

LD3 is a unique neural correlate of the team flow state, and LD5 is a unique neural correlate of a team (vs. no-team) state.

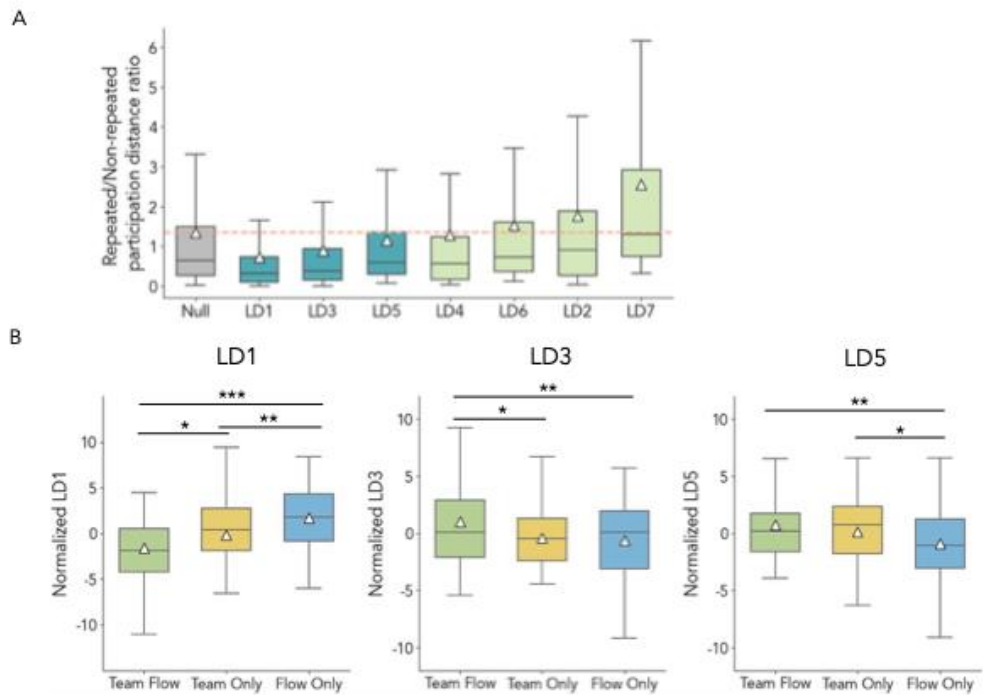

**Figure 4. Latent dimensions capture inter-individual and inter-condition differences.** (A) Inter-individual differences. Within repeated participation distance refers to the Euclidean distance between the two repetitions of one individual participant, between non-repeated participation distance refers to the Euclidean distance between any different participants. Boxplots shows the distribution of the ratio between the within repeated participation distance and between non-repeated participation distance for each LD (in an ascending order). Horizontal line shows the chance level. The ratio is significantly below chance level for LD1, LD3, LD5, suggesting that these latent dimensions contribute to the characterization of inter-individual differences. (B). Inter-condition differences. The latent dimension scores across three task conditions: team flow, flow only, and team only. LD scores were mean centered across conditions within each song. The horizontal line within each box represents the median, and the triangle represents the mean. Significant differences were found among conditions in LD1, LD3, LD5 using ANOVA. Asterisks denote significant post-hoc pairwise comparisons. *p<0.05, **p<0.01, ***p<0.001

**Latent dimensions were formed by distributed brain networks**

To understand the biological meaning of the latent dimensions, especially LD1, LD3, and LD5 that contributed the most to the inter-individual (trait) and inter-state differences, we traced the primary components contributing to them derived from the non-negative matrix factorization. Primary components of an LD were defined as components whose weights (on the LD) exceed the average + 3*standard deviation (i.e., as an empirical rule of selecting top 0.15% of the results) across all 160 NMF components. Based on this criterion, seven beta NMF components and two low-gamma NMF components were considered as primary components (Figure 5A): LD1 was featured by Beta NMF 1, Beta NMF 8, Gamma NMF 3, and Gamma NMF 30; LD3 was featured by Beta NMF 2, Beta NMF 3, and Gamma NMF 3; LD5 was featured by Beta NMF 1 and Beta NMF 11. Since we knew the linear transformation between NMF components and power spectral density channels, we further traced the contribution of each PSD channel in building up the primary NMF components (Figure 5B).

Alternatively, we can map each LD directly to the PSD channels by multiplying two weight matrices: a weight matrix from LD to NMF components (Extended Data Figure 5-1), and a weight matrix from NMF components to PSD channels (Extended Data Figure 5-2 – 5-8, column 1). We also highlighted the differences among the 3 task states for each LD by multiplying the LD to PSD map with the normalized PSD topographic map of each task state (Extended Data Figure 5-2 – 5-8, column 2-4).

In both approaches, we found that PSD from the beta and low-gamma frequency bands contained more valuable information than the theta and alpha frequency bands. However, within these frequency bands, we did not find a single brain region that primarily contributed to any LD. Instead, each LD was a weighted combination of many channels that are distributed across all the brain lobes, forming diverse network patterns.

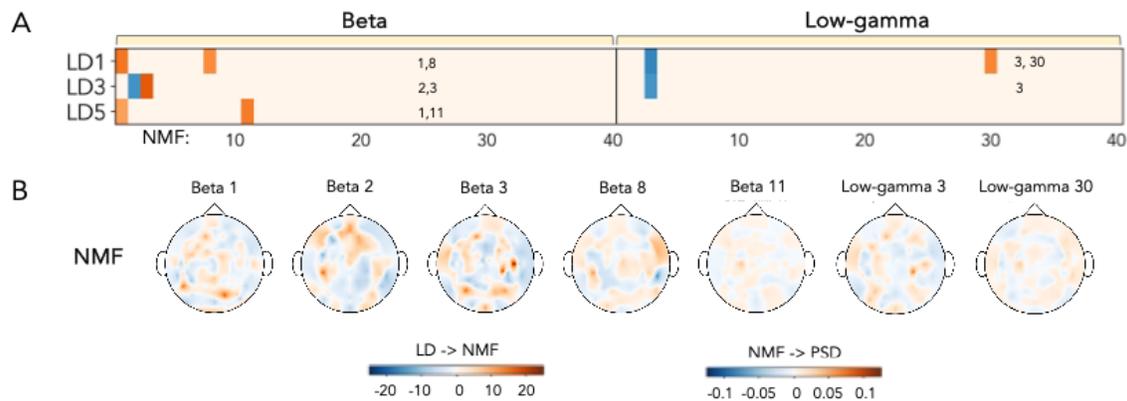

**Figure 5. Two-stage tracing of the major contributing EEG channels to the latent dimensions.** (A) The projection matrix from LD to NMF components. Primary NMF components whose weights in the projection matrix exceed 3 times standard deviations above/below the mean are highlighted. Only beta and low-gamma frequency bands are shown as there are no primary NMF components in the theta and alpha frequency bands. LD1 mainly consisted of beta NMF 1, beta NMF 8, low-gamma NMF 3, and low-gamma NMF 30; LD3 mainly consisted of beta NMF 2, beta NMF 3, and low-gamma NMF 3; LD3 mainly consisted of beta NMF 1, and beta NMF 11. Extended Data Figure 5-1 shows the full projection matrix. (B) The topographical map showing the weight of all 128 PSD channels on the primary NMF components. LD: latent dimension from the latent discriminant analysis, NMF: non-negative matrix factorization, PSD: power spectral

density. Extended Data Figure 5-2 ~ 5-8 show the topographical mapping of PSD to each LD.

**The latent EEG space shares a representational similarity pattern with a skill-cognition space.**

Finally, having understood the trait-state hierarchical structure of the latent space, we wonder whether the latent EEG traits and states are biologically meaningful - if they represent the functional differences of the brain (across individuals and across task conditions) instead of the brain anatomy alone, they should have some associations with behaviors. To address this question, we utilized representational similarity analysis (RSA) – a technique that helps establish connections between different spaces through their shared similarity structure (Kriegeskorte et al., 2008). Given the nature of the task, we constructed a skill-cognition space using seven behavioral trait and state measures: skill level (or performance), flow index, team index, song preference, and condition preference for each of the three conditions. Based on the Euclidean distance between pairs of samples in the skill-cognition space, we built a skill-cognition representational dissimilarity matrix (RDM) across all the experimental trials (Figure 6A). Similarly, we built a neural RDM from the latent EEG space as well (Figure 6B). We then assessed the correlation between the neural RDM and skill-cognition RDM, and found a significant correlation between them ($Pearson's\ r = 0.273, p < 0.001\ by\ permutation$), suggesting a shared similarity structure between the latent EEG space and skill-cognition space (Figure 6C-D, note that the association was also significant after excluding dissimilarities from different trials of the same participant, see Figure 6-1). In other words, if the neural characteristics of one individual during one trial was similar to that during another trial (either from the same individual or from a different individual), then the team flow experience during the game would be similar between the two trials.

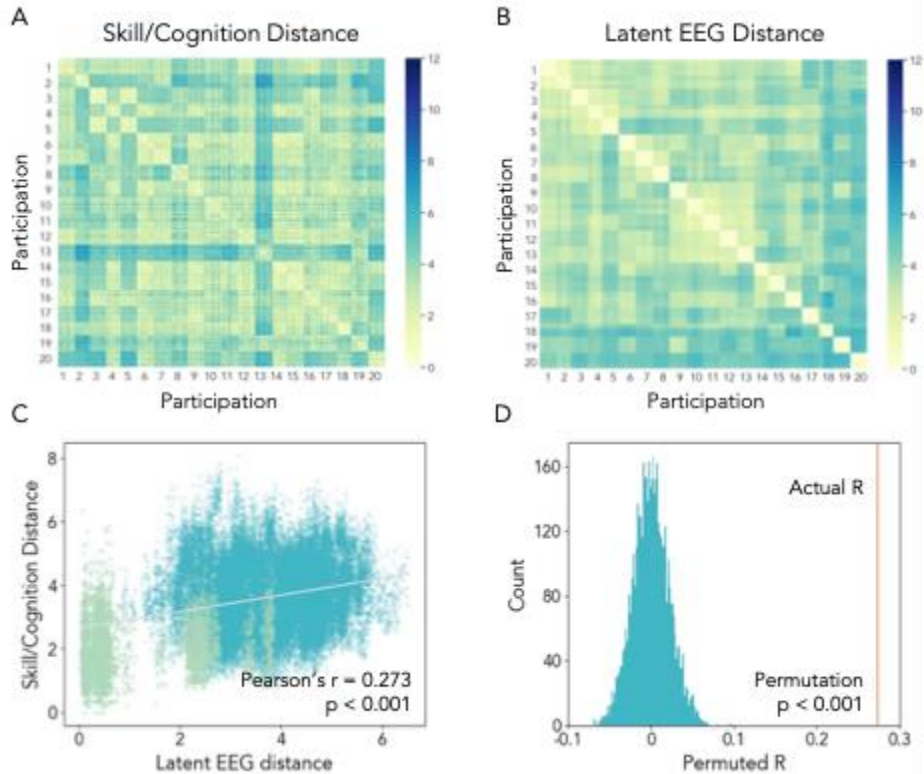

**Figure 6. Latent neural space shares a similarity pattern with a skill/cognition space.** Representational dissimilarity matrices (RDM) were built in (A) a skill/cognition space and (B) an latent EEG space across all trials. Each cell represent the Euclidean-distance between two trials. Trials from the same participation (i.e., one participant's one experiment session) are placed together, as indicated by participation labels 1-20. (C) Correlations between the RDM in the two spaces. Data from the same participant is highlighted in light green. Numbers indicate the Pearson's correlation and uncorrected p-value. (D) Permutation test of the RSA significance by calculating the correlation between RDMs for 5000 times after randomizing the column/row orders of the latent EEG RDM. A red line indicates the actual correlation, and the histogram indicates the correlation from the permutations. Numbers indicate the corrected p-value.

## Discussion

Our analysis pipeline offers several advantages over existing approaches: (1) it identifies both traits and states within a single neural latent space, (2) it provides evidence for a trait-state hierarchy in the neural space, (3) it maintains neurobiological interpretability, and (4) it correlates with a non-neural skill-cognition space that captures social interactions. However, our analysis has some limitations, including a small sample size, limited behavioral measurements, and restricted spatial resolution.

The first advantage of our identified latent space is its ability to classify individuals (traits) and task conditions (states) simultaneously in a unified framework. To our knowledge, existing EEG studies have only considered one aspect. Many EEG studies have investigated human neural states and built predictive models to classify certain states and cognitive processes, such as sleep stages (Supratak et al., 2017), emotions (Lin et al., 2010), and concentration (Edla et al., 2018). These studies focused on the generalizability of the models to predict the states, i.e., experimental conditions, without accounting for individual differences. Conversely, other studies focused solely on classifying or recognizing the identity of individuals based on EEG metrics – a technique known as 'EEG fingerprinting' (Jalaly Bidgoly et al., 2020; Rashid et al., 2020; Rosli et al., 2021). These fingerprinting studies did not account for different states within each individual. Therefore, to our knowledge, this is the first EEG-based analysis to model both traits and states in a single EEG latent space.

Second, our analysis pipeline provided empirical results for a trait-state dichotomy in the neural space, contributing new insights to existing psychological theories. Conceptually, traits systematically differ across individuals and remain stable across situations, while states vary within individuals depending on specific situations and moments in time. Mathematically, according to the whole trait theory, traits are density distributions of momentary states, and states oscillate around traits (Fleeson and Jayawickreme, 2015; Read et al., 2010). Our current findings revealed a novel hierarchical feature on top of the trait and state dichotomy. In our EEG latent space, we observed macro-segregations consisting of all the data from each participation, representing unique 'trait' clusters on the higher hierarchy, as well as micro-segregations consisting of the data from each task conditions within the trait clusters, suggesting smaller 'state' clusters that deviate from the center of their trait clusters. In the current dataset, a subset of participants participated twice in separate sessions; the finding that neural distances from repeated participations of the same participant were closer than those from different participants further emphasized the existence of trait-like features in the neural signals.

These findings highlight the importance of modeling individual differences in neural correlates of cognitive processes. Traditional approaches average EEG activities across participants under each task condition, treating individual differences as noise. In contrast, our results suggest that individual differences are a crucial component, as the same experimental manipulation may produce different effects across individuals. This trait-state hierarchy provides a valuable framework for future research.

Third, our pipeline preserves neurobiological interpretability. Existing EEG fingerprinting studies often prioritize algorithm performance, leveraging high-dimensional spaces and non-linear transformations at the cost of interpretability. In contrast, our approach goes beyond achieving classification goals by examining the contribution of each latent dimension (LD) to traits and states. Among the seven LDs, three (LD1, LD3, LD5) significantly explained variations in both individual traits and task states. LD1 differentiated all three states, LD3 separated team flow from non-team flow states, and LD5 distinguished team from non-team states. These results highlight how latent dimensions can reveal compound neural patterns that collaboratively distinguish traits and states, offering a data-driven perspective beyond traditional single-channel analyses.

Using linear transformations, we recovered the EEG channel weights for each LD and visualized their contributions. Consistent with prior findings (Shehata et al., 2021), beta and low-gamma frequency bands were more influential in the team flow task compared to alpha and theta bands. While several temporal lobe channels (e.g., NMF Beta 3, NMF Beta 8) aligned with previous studies, we found that LD neural correlates were widely distributed across the brain, including frontal, parietal, and occipital lobes. This supports the view of the brain as a distributed network system, where baseline network organization predicts traits (Dubois et al., 2018; Seitzman et al., 2019) and co-activation patterns predict task states (Cole et al., 2014). Social interaction engages multiple brain networks, including the salience, default mode, central executive, and subcortical networks (Feng et al., 2021). However, as existing evidence primarily comes from fMRI and PET studies, further research is needed to explore how our EEG-based findings translate to these broader contexts.

It is important to note that the macro-segregation may arise from various factors: it could reflect individual anatomical differences, such as brain volume, bone thickness, and 3D blood networks (Fiederer et al., 2016; Hagemann et al., 2008; Smit et al., 2012); functional differences, such as resting-state activity and its coherence network that establishes the baseline for neural information processing (Smit et al., 2008); or task-specific brain functions, such as responsiveness to musical rhythm. Using representational similarity analysis, we demonstrated a shared representation between the EEG latent space and a skill-cognition space, constructed from subjective surveys and performance data related to team flow during the task. This shared representation suggests that the latent EEG dimensions are predictive of the team flow experience. For example, if the EEG characteristics of an ongoing trial resemble those from a previous trial with high team flow, the participant is likely to experience high team flow again. Similarly, participants with closer proximity in the EEG space tend to report similar team flow experiences. These findings indicate that the macro-segregation in the latent space reflects functional differences relevant to social cognition and is not solely the result of anatomical differences.

In addition to its advantages, our study has several limitations. First, the sample size was small, constrained by COVID-19 restrictions during the experiment and the subsequent unavailability of professional players, which limited our ability to collect additional data. Larger datasets are needed to validate the robustness of our pipeline and findings across a broader range of participants and task types. Second, the number of available behavioral measures was limited. While we observed a close mapping between the latent EEG space and the skill-cognition space, the possibility remains that the latent EEG dimensions may correlate with other behavioral measures. Future research should incorporate more comprehensive behavioral assessments to explore additional neural-behavioral associations. Lastly, our analysis relies on the power spectral density of surface EEG channels, which offers limited spatial resolution. To improve the interpretation of latent dimensions, future studies could utilize alternative imaging modalities or EEG source localization techniques to achieve higher spatial precision.

## Conclusion

Using a two-stage dimensionality reduction pipeline combining non-negative matrix factorization and linear discriminant analysis, we extracted a seven-dimensional latent EEG space from the whole-brain power spectrum. This space revealed a hierarchical trait-state structure in the neural data, where individual participants formed primary clusters at the higher level, and task states within each participant formed sub-clusters. A subset of latent dimensions (LD1, LD3, LD5) significantly contributed to inter-individual and inter-state differences and could be linearly traced back to their primary EEG channels. Trial-to-trial similarities in this latent neural space corresponded to those in a skill-cognition space.

Our approach and findings have potential applications in broader contexts, including latent trait-state identification across different task states, dynamic trait-state-based team pairing, and predictive modeling of team behaviors.

## Materials and Methods

### Data source

We conducted all the analysis using a published team flow dataset (Shehata et al., 2021). In the study, 15 participants (5 males; age: 18-35 years) played a music rhythm game in 10 pairs (3 male pairs), among which 5 participants (1 male) were paired twice. We define a participation as a single experimental session for one participant, resulting in a total of 20 participations. Participants were screened and matched to pairs according to their skill level and song preference. During the task, while listening to a song, participants tapped a touch screen when the animated visual cues reached a designated area, receiving instantaneous positive feedback. The task consisted of three interleaved conditions: Team Flow (TF), Team Only (TO), and Flow Only (FO). Each condition contained 6 trials (1 song per trial). At the end of each trial, participants answered nine rating questions about their task experience, including their skill-demand balance (Q1 and Q2), feeling in control (Q3), automaticity (Q4), enjoyment (Q5), time perception (Q6), awareness of partner (Q7), teamwork (Q8), and coordination (Q9). Ratings on Q1-Q6 were averaged as an index of individual flow (flow index) (, and the ratings of Q7-Q9 were averaged as an index of team interaction (team index). The final dataset included 18 trials in total under three different task conditions. During the task, the EEG data of both players were collected simultaneously. Data was obtained with permission from Shehata et al (Shehata et al., 2021). More details about their data collection and processing can be found in their previous publication (Shehata et al., 2021).

### Power spectral density

Power spectral density (PSD) was estimated using Welch's overlapped segment averaging estimator. The PSD was first calculated for 20-25 epochs and then averaged within each trial, resulting in trial-wise PSD at each of the 128 channels. For each channel, the PSD was then averaged within delta (1 – 3 Hz), theta (4 – 7 Hz), alpha (8 -12 Hz), beta (13 – 30 Hz), and low-gamma (31 – 50 Hz) frequency bands. This analysis was conducted using the EEGLAB toolbox (Brunner et al., 2013; Welch, 1967) in MATLAB 2016a with default parameters, in the same way as described in the previous publication.

### Two-stage dimensionality reduction and latent space identification

To identify a latent EEG space that characterizes variabilities both across individuals and across task states, we conducted a two-stage dimensionality reduction on the PSD data, utilizing non-negative matrix factorization (NMF) and linear discriminant analysis (LDA).

In the first stage NMF dimensionality reduction, the input samples were participants' trial-wise data (total N=348), and the features were the PSD of all 128 EEG channels across 4 frequency bands (theta, alpha, beta, low-gamma, d=128*4=512). The goal of NMF was to reduce redundant information in the dataset and avoid overfitting during LDA (otherwise, the feature dimension would exceed the sample size, in which case the execution of the LDA algorithm is impossible). We normalized all the input data as follows: for each feature, we first normalized all three trials within each song (e.g., S1 TF, S1 FO, S1 TO) by subtracting the average PSD across these trials, in order to reduce the systematic noise due to song differences; next, we added the average PSD within each participation to their corresponding normalized trials to retain subject-specific PSD information in the data. To facilitate the interpretation of NMF components, we conducted NMF for each frequency band separately. During each NMF, the original data matrix **V** (348*128) was factorized into two non-negative matrices, a feature matrix **W** (348*p) and a coefficient matrix **H** (p*128), so that the multiplication of **W** and **H** can approximately reconstruct **V**. Here, p represents the number of NMF components/new features, each column in **W** is a new feature, and each column in **H** is the coefficients/weights of the new features to construct the original sample.

Next, we applied LDA to the output NMF components, including the top 40 NMF components for each frequency band (i.e., p=40 in each NMF model) as the new feature set. LDA is a supervised learning algorithm used for classification and dimensionality reduction (Fisher, 1936). It generates a set of latent dimensions (LD) through a linear combination of raw features. These latent dimensions maximize the separability of different classes by increasing the between-class distance and decreasing the within-class distance (Fisher, 1936). In our case, a class consisted of samples from one of the three conditions in one participant's one participation (e.g., participant 1's Team Flow trials when paired with participant 2, 5~6 trials in total). Participants who played twice with different partners were counted as two participations. The KFDA python package was used to perform LDA (Nikomborirak, n.d.).

**Determination of the number of dimensions**

In our pipeline, there were two hyperparameters to decide: the number of NMF components (n_NMF) from each frequency band that fed into the LDA, and the final number of LDs (n_LD) as the output. To make these decisions, we performed 5-fold cross-validation and used the final average validation accuracy as an objective criterion. Our goal was to find a trade-off between the dimensionalities and the generalizability of the classification algorithm. We repeated the cross-validation using different combinations of n_NMF and n_LD, with n_NMF ranging from 10 to 70 (step size = 5), and n_LD ranging from 1 to 20 (step size = 1), resulting in two accuracy heatmaps (Figure 2A). For each n_NMF, we calculated the area-under-curve (AUC) of the validation accuracy over all the n_LDs (Figure 2B). Note that the AUC here does not mean the AUC of an ROC curve, but represent an averaged accuracy across different choices of n_LD, thus its value can exceed 1. Because the accuracy AUC monotonically increased with the increase of n_NMF, we chose a local maxima of 40 as a balance point instead of the global maximum of 70. After deciding on the n_NMF, we examined how the corresponding validation accuracy (with the choice of n_NMF=40) changed with the increase of n_LD (Figure 2C). As it also increased monotonically but in a concave shape, we calculated the location of a knee (or elbow) point on this curve, at which the increase of accuracy along the n_LD was no longer rapid (i.e., the benefit of including one more LD became small). This n_LD was found to be 7. The knee point detection was conducted using the 'kneed' package in Python (Satopaa et al., 2011).

**Control analysis of the LDA classification performance**

To rule out the possibility that the data clustering pattern in the latent space resulting from LDA is an overfitting of noise, we conducted three types of permutation test: 1) randomly shuffling both the participation and task condition labels, 2) randomly shuffling the task condition labels while keeping the participation labels the same, and 3) randomly shuffling the participation labels while keeping the task condition labels the same. For each type of permutation, we repeated the same procedure as in the initial identification of the latent space – we selected the top 40 NMF components per frequency band, and conducted LDA with 5-fold cross-validations to assess the average training and validation accuracy. Each permutation was repeated 500 times, so that the distribution of the validation accuracy served as a null distribution and was further compared with the validation accuracy of the original dataset (Figure 2-1).

**Confirmatory analyses for the macro- and micro-segregation model**
**Test of the latent space architecture:** Our goal was to test among three alternative hypotheses about the hierarchical structure of the latent EEG space: 1) state-trait hierarchy, 2) trait-state hierarchy, and 3) no hierarchy. To achieve this goal, we defined inter-individual distances and intra-individual distances. We first calculated the centroid (i.e., the average 7D coordinates of all the trial samples) of each task condition from each participation/individual. Then we defined an inter-individual (or intra-state) distance as the Euclidean distance between the centroids of two participations/individuals with the same task condition (i.e., $P_xS_a - P_yS_a$), and an intra-individual (or inter-state) distance as the Euclidean distance between two task conditions for the same participation/individual (i.e., $P_xS_a - P_xS_b$, Figure 3A-C). The inter/intra-individual distance ratio was calculated as the mean inter-individual distance divided by the mean intra-individual distance. We also constructed a null distribution of the inter/intra-individual distance ratio by performing the aforementioned calculations over 5000 random permutations of participation and task condition labels. Finally, we compared the inter/intra individual distance to the null distribution: if the actual distance is not significantly different from the null distribution, there is no hierarchy in the data; if the actual distance is significantly above the null distribution, there is a trait-state hierarchy; and if the actual distance is significantly below the null distribution, there is a state-trait hierarchy. We calculated the p-value as the percentile of the actual distance on the null distribution of the distance.
**Hierarchical clustering:** We performed a hierarchical clustering in the latent neural space among all participations of all participants. Euclidean distances between participation centroids (i.e., average 7D coordinates of all the trial samples within a participation) were calculated. Based on these distances, participations that are most proximate to each other were paired into binary clusters, and these binary clusters were grouped into larger clusters given the distances between binary clusters (Figure 3E).

**Representational similarity analysis**
We tested whether the identified EEG latent space shared a similarity structure with a non-neural, psychological space using representational similarity analysis (RSA). We first constructed a skill-cognition space using seven variables: flow index, team index, skill level, song preference, and condition preference for each task condition. For both the EEG latent space and the skill-cognition space, we then calculated the Euclidean distance between every pair of trials. These distances served as indices of dissimilarity, producing two dissimilarity matrices (Figure 4A, 4B). Next, we calculated Pearson's correlation between the dissimilarities of EEG latent space and the skill-cognition space and tested the significance of correlation through permutation (Figure 4C, 4D). During the permutation test, we shuffled the rows/columns of the skill-cognition dissimilarity matrix and calculated a new correlation coefficient between the two dissimilarity matrices. We repeated this procedure 5000 times, yielding a null distribution of the correlations. We finally obtained a significance level by comparing the actual correlation with the null distribution.

**Quantification of the LD contribution to inter-individual and inter-state differences**
For inter-individual differences, we derived a measure to quantify the ability of a latent dimension to minimize the distance between the repeated participations of the same individual, while maximizing the distance between different individuals: repeated vs. non-repeated participation distance ratio. For each LD, we calculated all the distances between the centroids of two participations from any repeated participant as a set of within-repeated-participation-distance, as well as all the distances between the centroids of two different participants as a set of between-non-repeated-participation distance. This allowed us to generate a distribution of the ratio between these two metrics. Meanwhile, we performed bootstrapping 1000 times by repeatedly sampling a within-repeated-participation distance (of one LD) and a between-non-repeated-participation distance (of another LD, may or may not be the same as the previous one) to generate a null distribution of the ratio. Finally, we conducted two-sample t-tests between the ratio distribution of every LD and the null distribution to identify LDs that had a ratio significantly smaller than the null distribution. All the reported p-values were after Bonferroni correction of multiple comparisons.

For inter-state differences, we first normalized the data by subtracting the average LD value across three trials within each song, and then performed one-way ANOVA on each LD across the three task states (TF, FO, TO), as well as the post-hoc Tukey's t tests between every pair of states. LDs that were significantly different across the three task states were considered to contribute to the characterization of inter-state differences.

## Code and Software

Analysis code for the manuscript is publicly available at
https://github.com/wuqy052/team_flow_latent_trait_state .

## Acknowledgements

This work has been supported by Japan Society for Promotion of Science (JSPS), Grants-in-Aid for Scientific Research (Fostering Joint International Research(B), Grant Number 18KK0280 for MS and SN. QW was supported by the Caltech NIMH Conte Center (P50MH094258) and Tianqiao and Chrissy Chen Graduate Fellowship. We thank Ralph Adolphs and Yue Xu for a helpful discussion and feedback.

## Conflict of Interest

Authors report no conflict of interest.

# Extended Data Figure

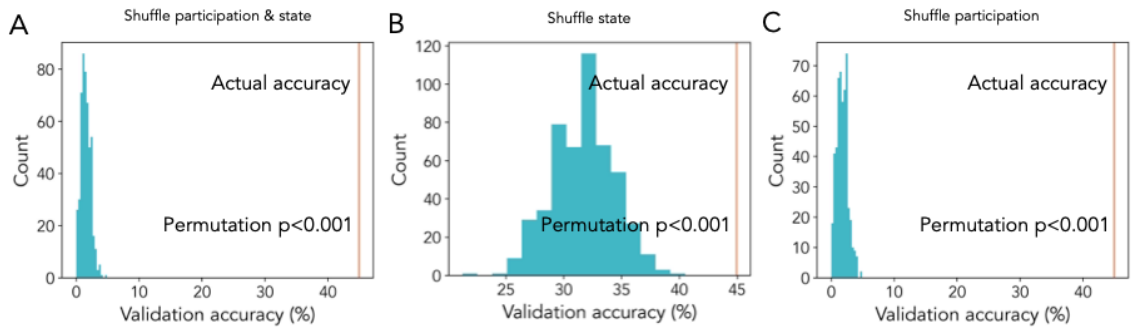

**Extended Data Figure 2-1. Permutation tests on the pipeline's validation accuracy.** Permutation-based null distribution (histogram) and the actual accuracy from our pipeline (vertical line). Three types of permutation tests were conducted, by (A) shuffling both participation and task state labels, (B) shuffling only the task state label, and (C) shuffling only the participation label, respectively.

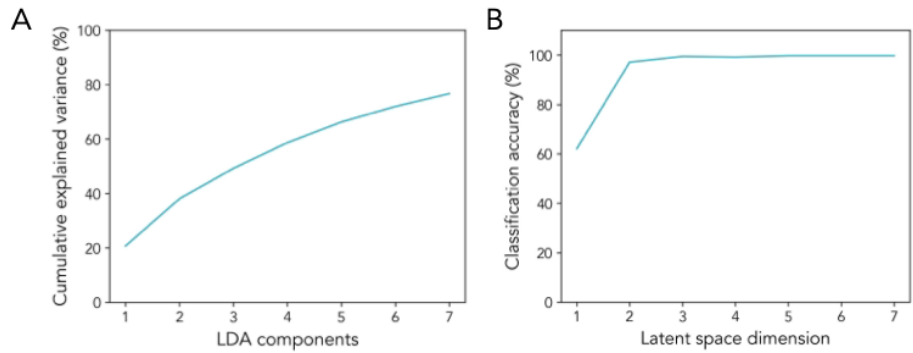

**Extended Data Figure 2-2. Model evaluation.** (A) Cumulative explained variance by number of LDs. The 7-dimensional LD space can explain 76.64% between vs. within group variance. (B) In-sample classification accuracy. The current latent space can classify the participation*condition classes at >99% accuracy with at least 3 latent dimensions.

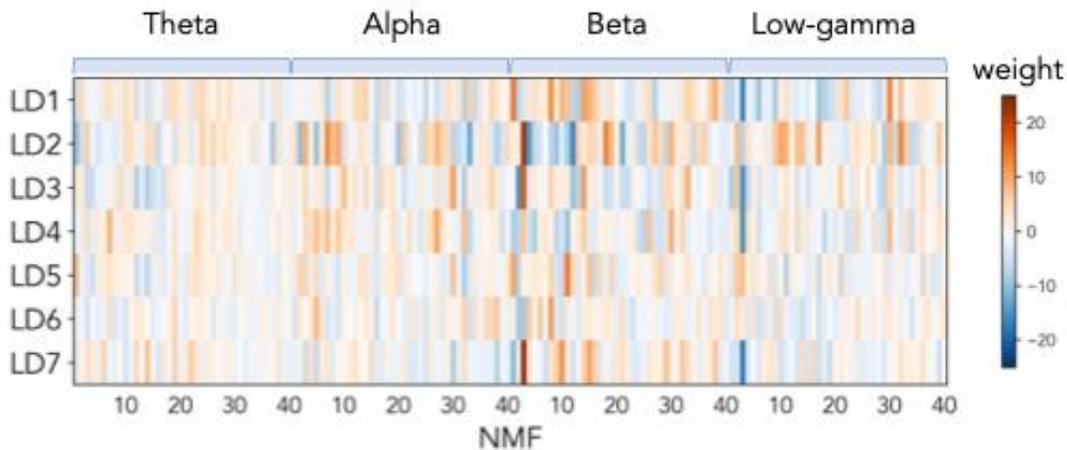

**Extended Data Figure 5-1. Full LDA to NMF projection matrix.** Each row represents the weight of all the 160 NMF components (40 per frequency band) when constructing an LDA latent dimension. Larger magnitude indicates higher contribution.

(A)

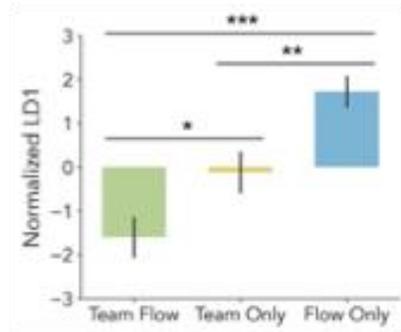

(B)

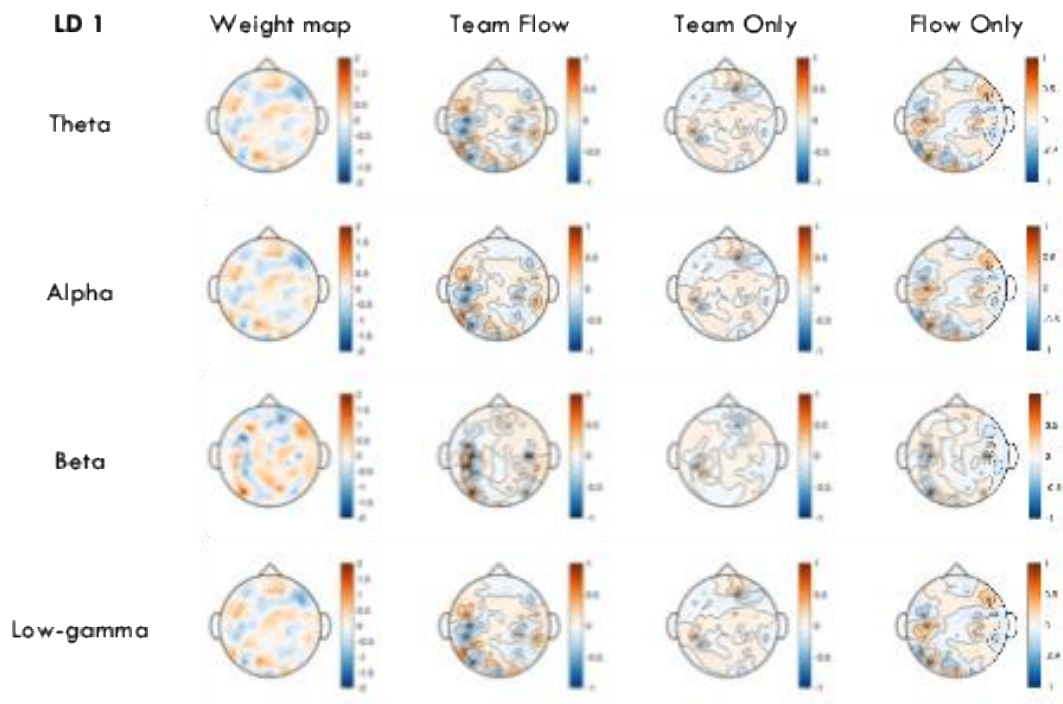

**Extended Data Figure 5-2. Comparison of the LD 1 across three task states.** (A) Mean-normalized LD 1 under three task states: Team Flow, Flow Only and Team Only. Error bars indicate the standard error. These three states had significantly different LD values, asterisks show the significance level of the post-hoc pairwise Tukey's test. *p<0.05,**p<0.01,***p<0.001. (B) Topographical map of the weight of each PSD channel to LD 1 (column 1), and the topographical map of weighted PSD value under three task states (column 2-4, i.e., multiplication of the normalized PSD per channel and the PSD to LD weight per channel).

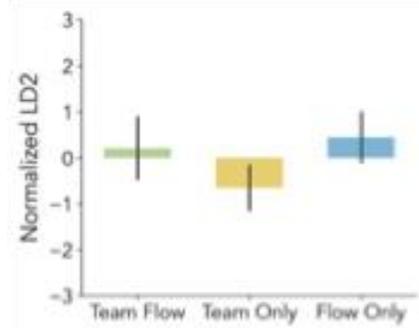

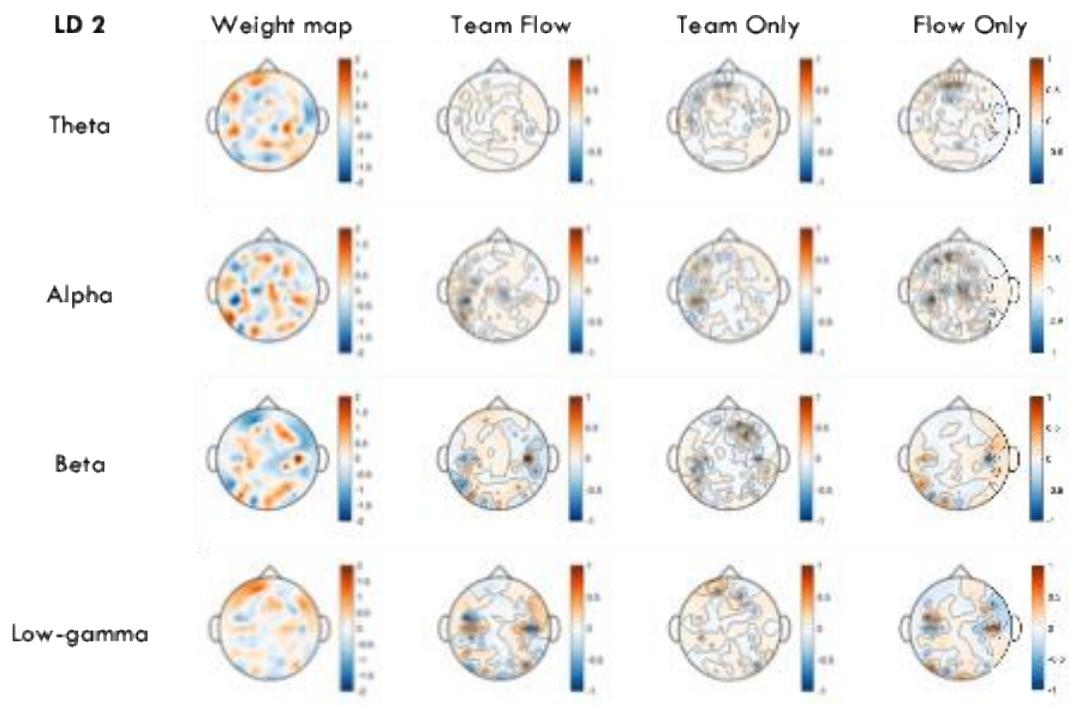

**Extended Data Figure 5-3. Comparison of the LD 2 across three task states.** (A) Mean-normalized LD 2 under three task states: Team Flow, Flow Only and Team Only. Error bars indicate the standard error. LD 2 across three states were not significantly different from each other. (B) Topographical map of the weight of each PSD channel to LD 2 (column 1), and the topographical map of weighted PSD value under three task states (column 2-4, i.e., multiplication of the normalized PSD per channel and the PSD to LD weight per channel) .

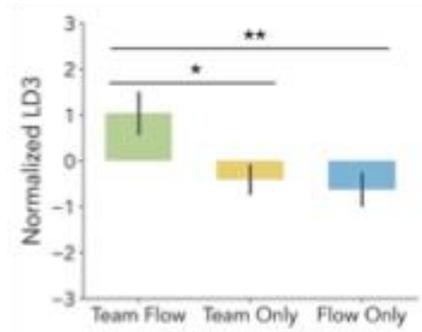

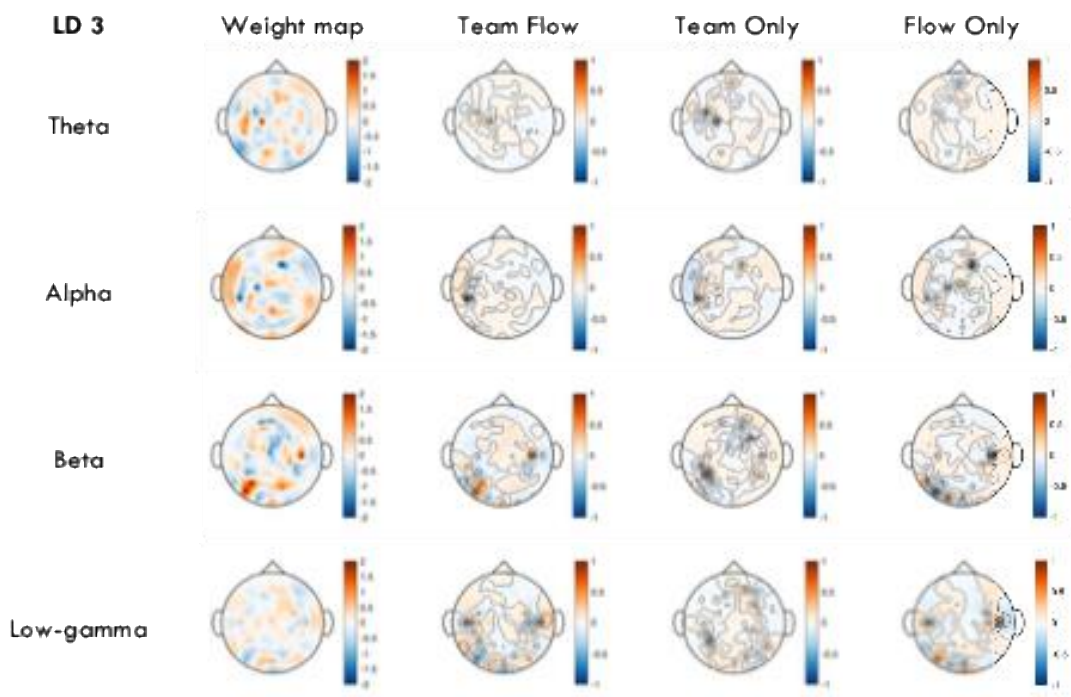

**Extended Data Figure 5-4. Comparison of the LD 3 across three task states.** (A) Mean-normalized LD 3 under three task states: Team Flow, Flow Only and Team Only. Error bars indicate the standard error. These three states had significantly different LD values, asterisks show the significance level of the post-hoc pairwise Tukey's test. *p<0.05,**p<0.01,***p<0.001. (B) Topographical map of the weight of each PSD channel to LD 3 (column 1), and the topographical map of weighted PSD value under three task

states (column 2-4, i.e., multiplication of the normalized PSD per channel and the PSD to LD weight per channel).

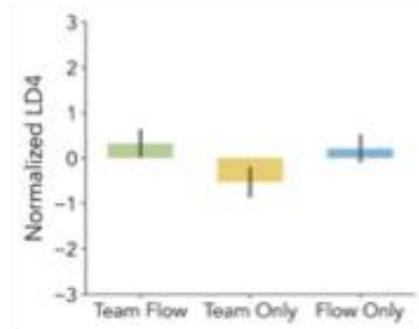

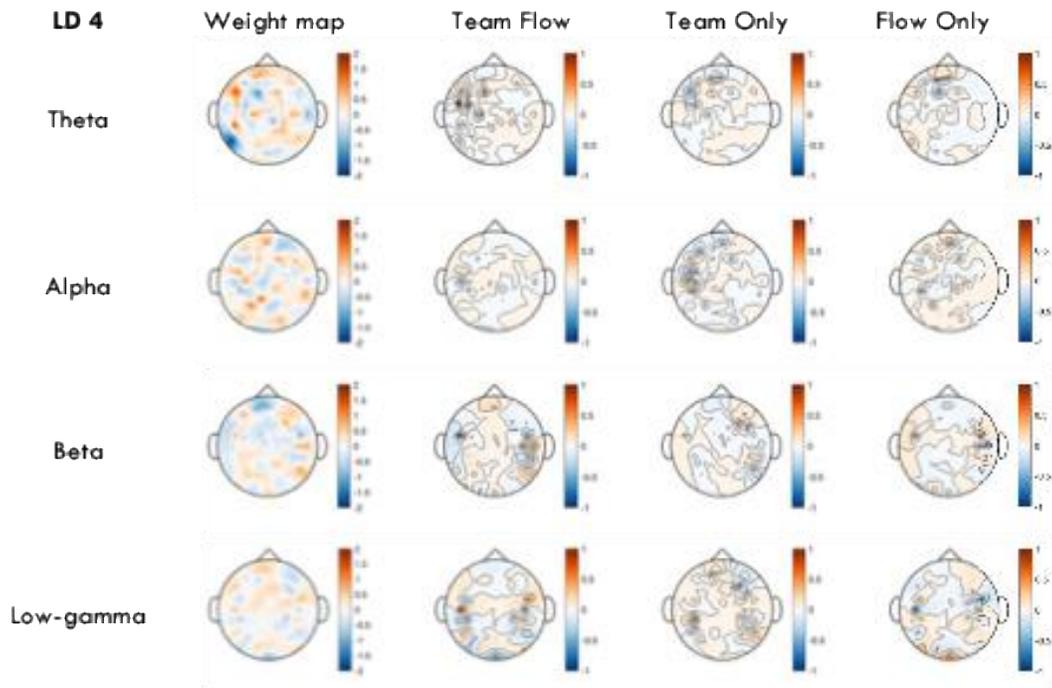

**Extended Data Figure 5-5. Comparison of the LD 4 across three task states.** (A) Mean-normalized LD 4 under three task states: Team Flow, Flow Only and Team Only. Error bars indicate the standard error. LD 4 across three states were not significantly different from each other. (B) Topographical map of the weight of each PSD channel to LD 4 (column 1), and the topographical map of weighted PSD value under three task states

(column 2-4, i.e., multiplication of the normalized PSD per channel and the PSD to LD weight per channel).

(A)

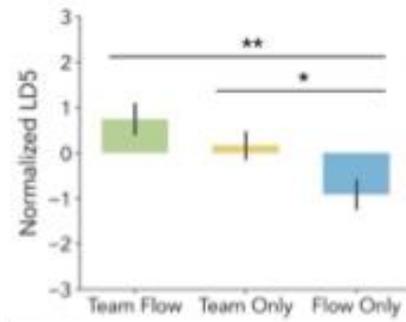

(B)

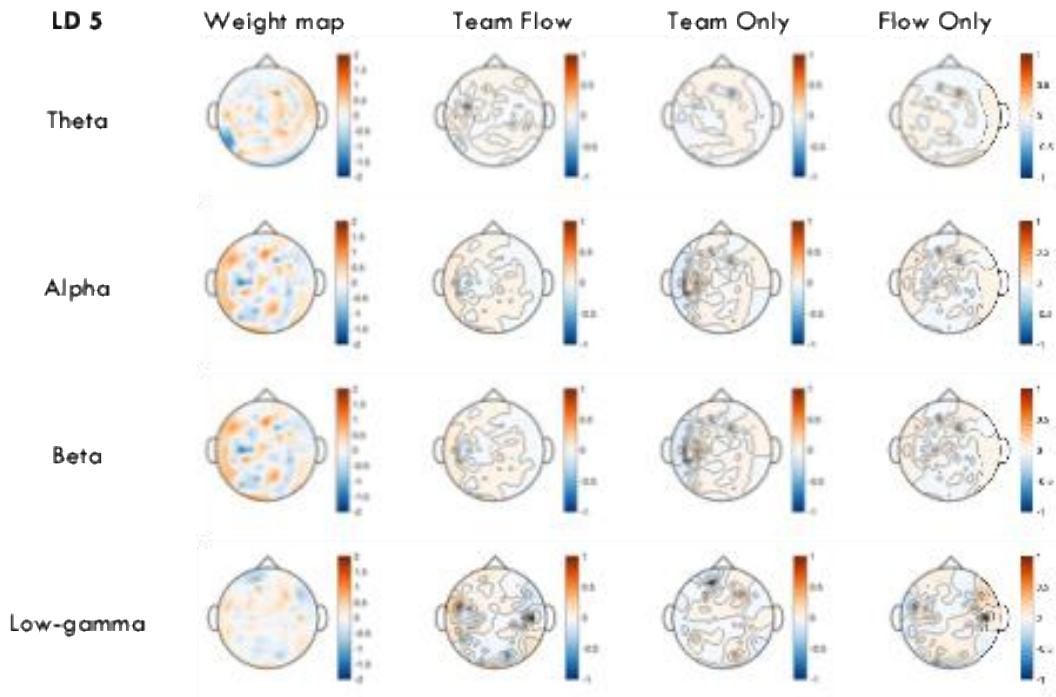

**Extended Data Figure 5-6. Comparison of the LD 5 across three task states.** (A) Mean-normalized LD 5 under three task states: Team Flow, Flow Only and Team Only. Error bars indicate the standard error. These three states had significantly different LD values, asterisks show the significance level of the post-hoc pairwise Tukey's test. *p<0.05,**p<0.01,***p<0.001. (B) Topographical map of the weight of each PSD channel to LD 5 (column 1), and the topographical map of weighted PSD value under three task

states (column 2-4, i.e., multiplication of the normalized PSD per channel and the PSD to LD weight per channel).

(A)
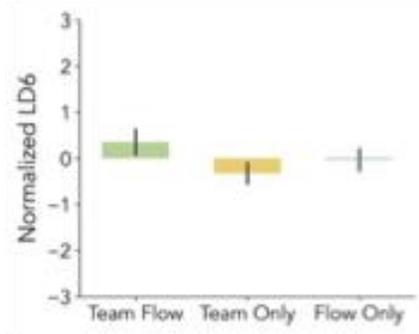

(B)
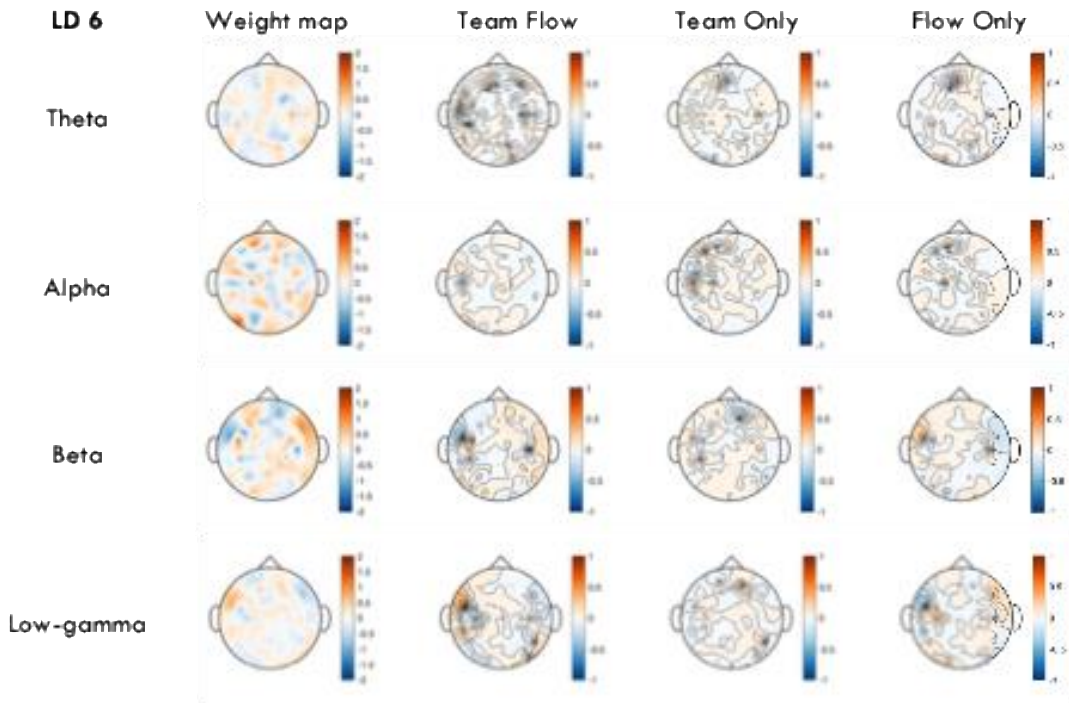

**Extended Data Figure 5-7. Comparison of the LD 6 across three task states.** (A) Mean-normalized LD 6 under three task states: Team Flow, Flow Only and Team Only. Error bars indicate the standard error. LD 6 across three states were not significantly different from each other. (B) Topographical map of the weight of each PSD channel to LD 6 (column 1), and the topographical map of weighted PSD value under three task states

(column 2-4, i.e., multiplication of the normalized PSD per channel and the PSD to LD weight per channel).

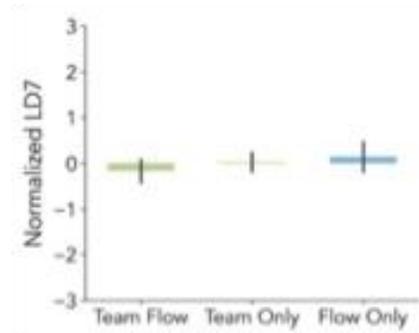

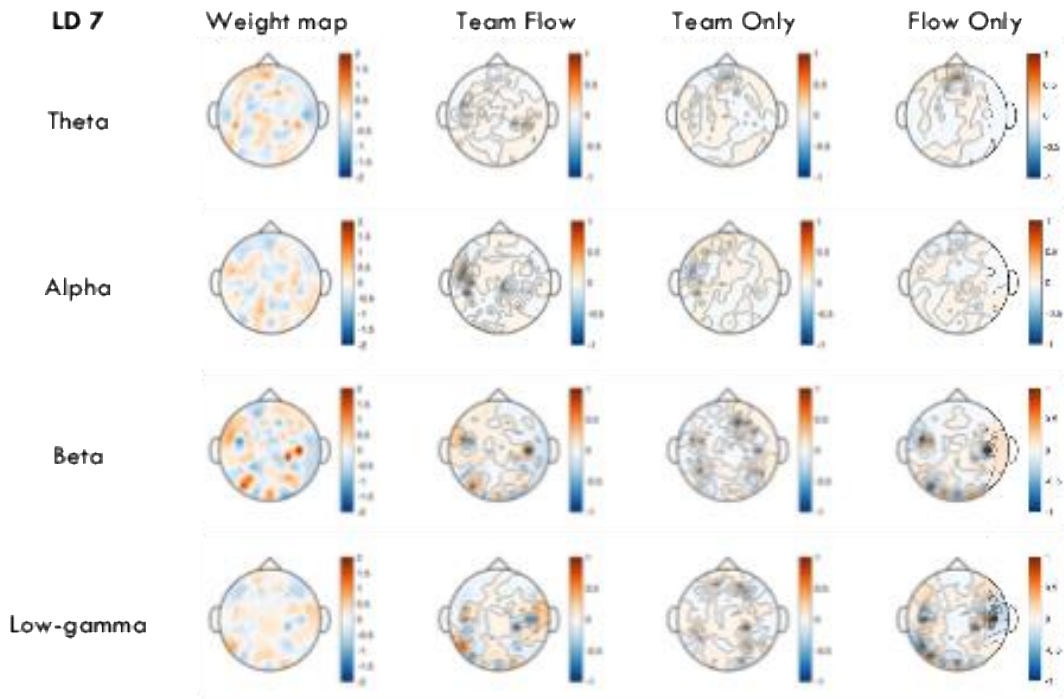

**Extended Data Figure 5-8. Comparison of the LD 7 across three task states.** (A) Mean-normalized LD 7 under three task states: Team Flow, Flow Only and Team Only. Error bars indicate the standard error. LD 7 across three states were not significantly different from each other. (B) Topographical map of the weight of each PSD channel to LD 7 (column 1), and the topographical map of weighted PSD value under three task states

(column 2-4, i.e., multiplication of the normalized PSD per channel and the PSD to LD weight per channel).

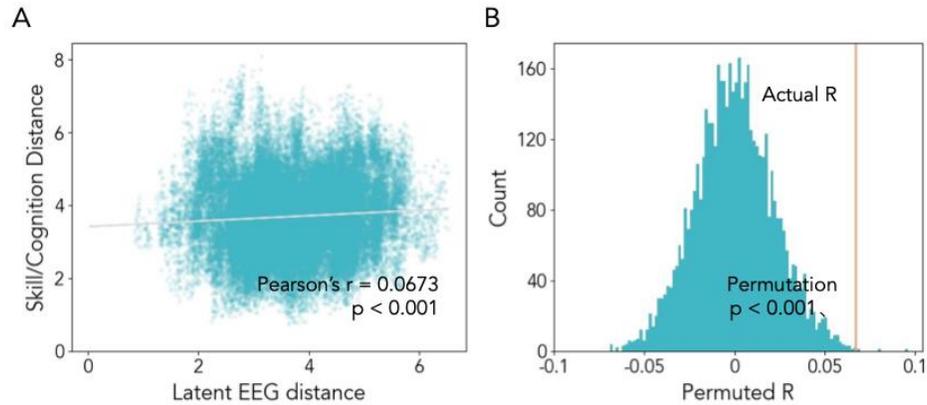

**Extended Data Figure 6-1. RSA without same-participant data.** (A) Correlations between the RDM in the two spaces. Data from the same participant is removed from Figure 4C. Numbers indicate the Pearson's correlation and uncorrected p-value. (D) Permutation test of the RSA significance by calculating the correlation between RDMs for 5000 times after randomizing the column/row orders of the latent EEG RDM. A red line indicates the actual correlation, and the histogram indicates the correlation from the permutations. Numbers indicate the corrected p-value.